\documentclass[3p,times,twocolumn]{elsarticle}

\usepackage{ecrc}

\volume{00}

\firstpage{1}

\journalname{Nuclear and Particle Physics Proceedings}

\runauth{}

\jid{nppp}

\jnltitlelogo{Nuclear and Particle Physics Proceedings}

\usepackage{amssymb}

\biboptions{sort&compress}

\usepackage[figuresright]{rotating}

\usepackage{tabularx}
\usepackage{multirow}

\begin{document}

\begin{frontmatter}



\dochead{}

\title{Supersymmetry, direct and indirect constraints}


\author{Farvah Mahmoudi}

\address{Univ. Lyon, Univ. Lyon 1, CNRS/IN2P3, Institut de Physique Nucl\'eaire de Lyon,\\ UMR5822, F-69622 Villeurbanne, France}

\address{Theoretical Physics Department, CERN, CH-1211 Geneva 23, Switzerland}

\begin{abstract}

We present an overview of direct and indirect constraints in the MSSM, in CP-conserving and CP-violating MSSM scenarios, with some emphasis on the importance of combining the constraints from different sectors, namely  SUSY and Higgs direct searches at the LHC, flavour physics, dark matter and electric dipole moments.
\end{abstract}


\end{frontmatter}


\section{Introduction}
\label{}

The Minimal Supersymmetric extension of the Standard Model (MSSM) is a well motivated and extensively studied scenario beyond the Standard Model (SM). It is a prototypical UV-complete model, for which many dedicated tools have been developed. With more than one hundred parameters, the MSSM is however difficult to explore in a systematic way. For this reason, most of the studies have been performed in constrained scenarios assuming specific SUSY breaking mechanisms with only a handful number of parameters. In absence of New Physics (NP) signals at the LHC, more systematic studies in general MSSM scenarios have emerged \cite{Berger:2008cq,AbdusSalam:2009qd,Sekmen:2011cz,Arbey:2011un,Arbey:2011aa}. In the following we present highlights on the current direct and indirect constraints in the phenomenological MSSM (pMSSM), which is the most general CP and R-parity conserving scenario assuming minimal flavour violation, described by 19 parameters. The CP-violating version of the pMSSM which differs from the CP-conserving case by the addition of six independent CP-violating phases will also be considered.

\section{CP-conserving MSSM}

To study the CP-conserving pMSSM, we perform random scans in the parameter space using SOFTSUSY~\cite{Allanach:2001kg}, varying the SUSY masses between 0 and 3 TeV, the trilinear couplings between -10 and +10 TeV, and $\tan\beta$ between 2 and 60, and assuming the neutralino 1 to be the lightest supersymmetric particle, so that it constitutes a dark matter candidate.

\subsection{Direct searches}

Direct searches at the LHC are persued in many different channels, and in particular squark and gluino direct searches (jets $+ E_T\!\!\!\!\!/$ ), stop and sbottom direct searches ($t$, $b$-jets ($+$ leptons) $+ E_T\!\!\!\!\!/$ ) and chargino and neutralino direct searches (leptons ($+ b$-jets) $+ E_T\!\!\!\!\!/$ ) at ATLAS and CMS which are considered in this study for both Run 1 and Run 2. To this end, events are generated with MadGraph~\cite{Alwall:2014hca} and/or Pythia~\cite{Sjostrand:2007gs}, and the detector response is simulated with Delphes~\cite{deFavereau:2013fsa} (see Refs.~\cite{Arbey:2011un,Arbey:2015hca} for a description of the employed tools and methodology). The events are then compared with the published backgrounds to determine whether a parameter point is excluded. In addition to the SUSY searches, we consider mono-X searches. Monojets searches provide the strongest constraints, and correspond to the search for 1 hard jet + $E_T\!\!\!\!\!/$ , which is often considered as the dark matter search at the LHC. However, in the MSSM one needs to recast the results as monojets can be constituted of one hard jet plus soft jets and $E_T\!\!\!\!\!/~$~\cite{Arbey:2015hca,Arbey:2013iza}. Therefore, monojet searches are particularly constraining in the MSSM in the cases where the strongly interacting sparticles (squarks, gluino) have a small mass splitting with the lightest neutralino, whose decays generate soft jets.

\begin{figure}[!t]
\begin{center}
\includegraphics[width=6.5cm]{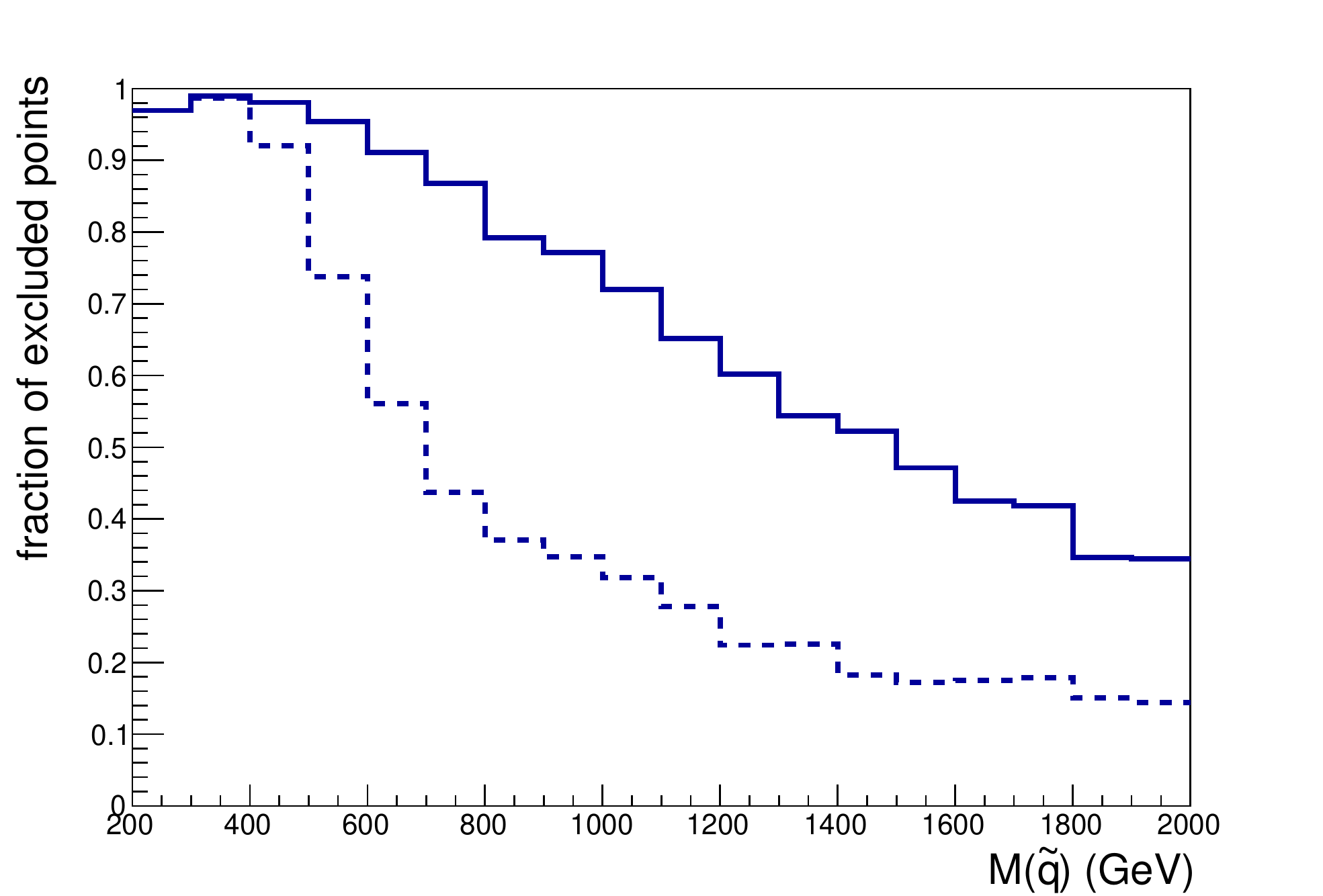}\\ \includegraphics[width=6.5cm]{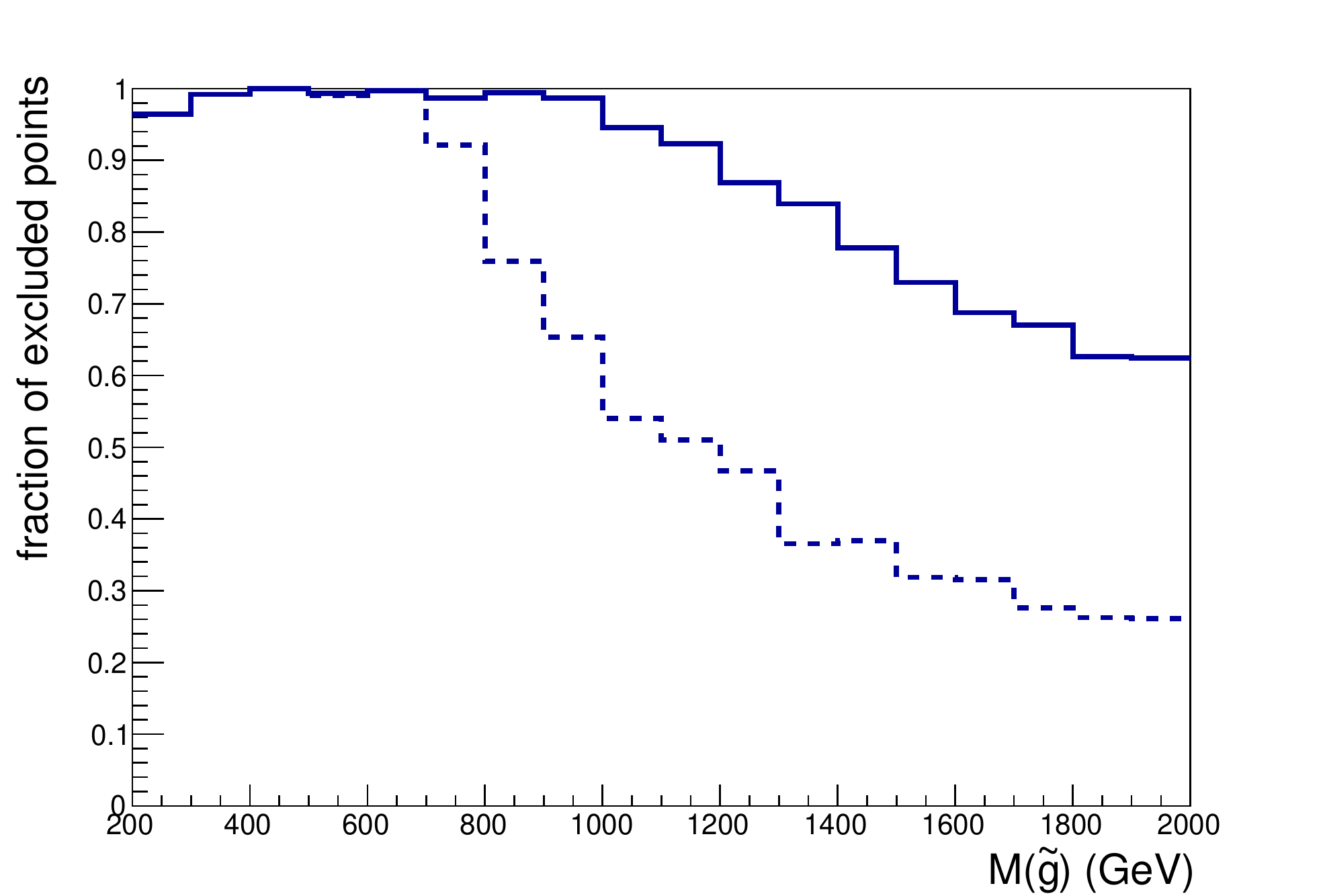}
\caption{Fraction of excluded points as a function of the lightest squark mass (upper panel) and gluino mass (lower panel). The dotted lines correspond to 8 TeV results, and the solid lines to 8+13 TeV. \label{fig:LHCdirect}}
\end{center}
\end{figure}

In Figure \ref{fig:LHCdirect}, we show the fraction of pMSSM points excluded by the LHC direct searches at 8 and 13 TeV, as a function of the lightest of the first and second generation squarks and of the gluino mass. First, we observe that squarks above 500 GeV can easily escape detection. In addition, there are still a few parameter points where light squarks below 500 GeV can escape. Concerning the gluinos, most of them are excluded with masses below 1 TeV. This result sharply contrasts with the results obtained for simplified and constrained SUSY scenarios, where squarks and gluinos below 2-3 TeV are not viable. This can be understood as it is possible in the pMSSM to have long decay chains in compressed or complicated scenarios.

\subsection{Indirect constraints}

Indirect constraints can also set strong limits on the MSSM. In particular, the mass of the Higgs boson as well as the measurements of its couplings at the LHC, set strong constraints on the Higgs and stop sectors of the MSSM \cite{Arbey:2011ab,Arbey:2012dq,Arbey:2012bp}. Similarly, flavour physics observables can lead to strong limits on the MSSM parameter space. In particular, the branching ratio of $B_s \to \mu^+ \mu^-$, which has been measured by the LHCb and CMS Collaborations~\cite{CMSandLHCbCollaborations:2013pla}, is very sensitive to the mass of the pseudo-scalar Higgs boson at large $\tan\beta$ \cite{Arbey:2012ax}, and the inclusive branching fraction of $B \to X_s \gamma$ is sensitive to the charged Higgs boson as well as the stops and charginos \cite{Mahmoudi:2007vz}. These observables are complementary to the direct searches for heavy Higgs bosons \cite{Arbey:2013jla}.

\begin{figure}[!t]
\begin{center}
\includegraphics[width=7.5cm]{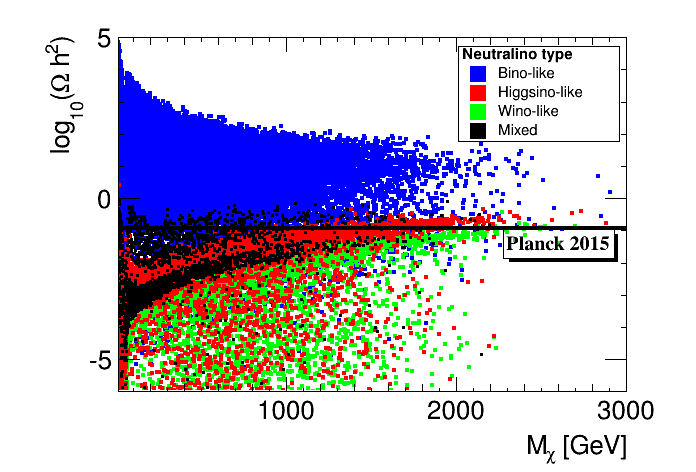}\\
\includegraphics[width=7.5cm]{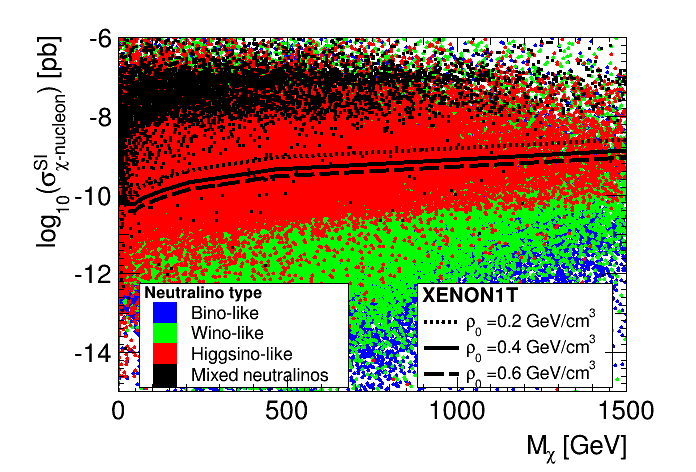}
\caption{Neutralino relic density (upper pannel) and scattering cross section with proton (lower pannel) as a function of the neutralino 1 mass and type. The lines correspond to the observational and experimental limits of the Planck and XENON1T collaborations. \label{fig:dmobs}}
\end{center}
\end{figure}

Further constraints can be set on the MSSM assuming that the lightest neutralino constitutes dark matter. First, the relic density can be computed assuming that it is a thermal relic \cite{Arbey:2009gu} and compared to the Planck limit~\cite{Aghanim:2018eyx}. Second, dark matter direct detection experiments such as XENON1T~\cite{Aprile:2017iyp} can constrain the scattering cross section of neutralino 1 with nucleons. Third, indirect detection can set constraints on the annihilation cross sections of neutralinos into SM particles. Even if the dark matter sector constraints suffers from uncertainties \cite{Baro:2007em,Arbey:2017eos,Robbins:2017wgv,Arbey:2018uho}, they can strongly constrain the pMSSM parameter space, and are very sensitive to the nature of the neutralino 1, as can be seen in Figure \ref{fig:dmobs}. 

\begin{figure}[!t]
\begin{center}
\includegraphics[width=7cm]{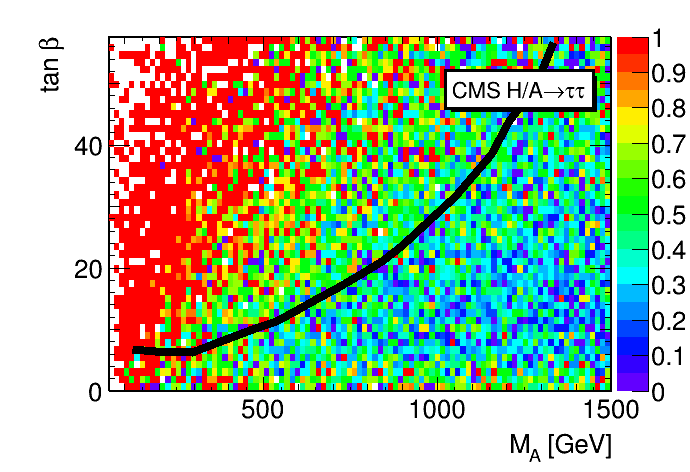}
\caption{Fraction of points excluded by flavour physics and dark matter direct detection in the $(M_A,\tan\beta)$ parameter plane.  The line delimits the region excluded by the heavy Higgs $H/A \to \tau\tau$ searches at the LHC.\label{fig:matanb}}
\vspace*{0.2cm}\includegraphics[width=7cm]{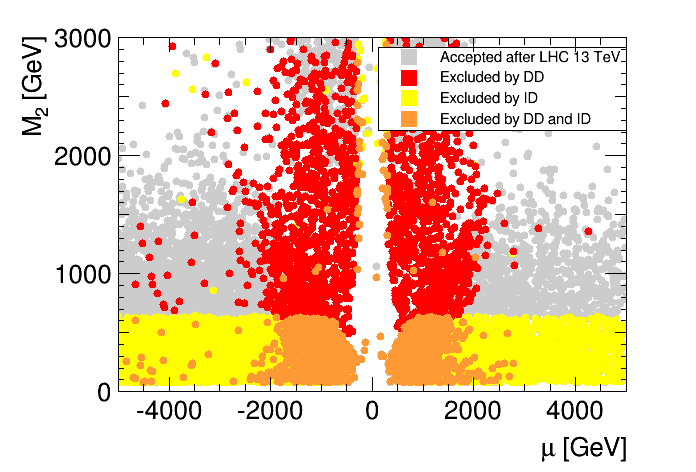}
\caption{In the $(\mu,M_2)$ parameter plane, points excluded by the LHC Higgs and SUSY searches (in gray), as well as dark matter direct detection (in red), indirect detection (in yellow), and both types of detections (in orange).\label{fig:m2mu}}
\end{center}
\end{figure}

The complementarity of the constraints from different sectors is of utmost importance, as it will constitute the only way to identify the underlying theory in case of discovery of new phenomena or particles at colliders or in space \cite{Arbey:2017eos,Arbey:2015aca}. Figures \ref{fig:matanb} and \ref{fig:m2mu} illustrate the complementary between different searches, showing that Higgs, flavour, dark matter and supersymmetric particle searches probe different regions of the pMSSM parameter space, and that in case of detection of new particles, only the combination of all these constraints will allow us to disentangle the parameters of the underlying theory.

\section{CP-violating MSSM}

So far, we investigated the observables which are only weakly sensitive to CP-violation, so that the pMSSM was well suited. However, the MSSM can contain many possible sources of CP violation beyond the SM.
We therefore extend our analysis by adding to the standard pMSSM 6 CP-violating phases, corresponding to the phases of the $M_{1,2,3}$ masses and $A_{t,b,\tau}$ trilinear couplings, which is the minimal extension of the pMSSM to include CP-violation, for a total of 25 parameters. The CP phases can take values between $-180$ and 180 degrees, and modify the mixing matrices and couplings \cite{Lee:2003nta}. The main phenomenological difference with the CP-conserving pMSSM is that the three neutral Higgs bosons are now mixed, giving three states $h_1$, $h_2$, $h_3$ with scalar and pseudoscalar components.

\subsection{Electric dipole moment constraints}

Electric dipole moments (EDM) are the most sensitive observables to CP-violation. In absence of CP-violation, the EDMs are extremely small, hence any deviation would constitute a proof for the existence of NP. Let us consider the following Lagrangian density:
\begin{equation}
 \mathcal{L}_{\mbox{EDM}}= - \frac{i}{2} d_f F^{\mu\nu} \bar{f} \sigma_{\mu\nu} \gamma_5 f \,,
\end{equation}
where $f$ corresponds to the SM fermions and $d_f$ their corresponding EDM. The quark EDMs are however not observed and only the nucleon EDM can be seen which is related to the quark EDMs by:
\begin{equation}
 d_N = \eta^E(\Delta^N_d d_d + \Delta^N_u d_u + \Delta^N_s d_s)\,,
\end{equation}
where $\Delta^N_q$ and $\eta^E$ are of order 1.

The current experimental limits at 95\% C.L. are given in Table 1. 
\begin{table}[h!]
\scalebox{0.8}{\small\begin{tabular}{|c|c|c|}
 \hline
 EDM & Upper limit (e.cm) & Equivalent limit (e.cm) \\
 \hline\hline
 Thallium \cite{Regan:2002ta} & $1.3\times10^{-24}$ & $|d_e|: 2.1\times10^{-27}$ \\
 \hline
 Thorium monoxide  \cite{Baron:2013eja} & - & $|d_e|: 1.1\times10^{-28}$ \\
 \hline\hline
 Muon \cite{Bennett:2008dy} & $1.9\times10^{-19}$ & $|d_\mu|: 1.9\times10^{-19}$\\
 \hline\hline
 \multirow{2}{*}{Mercury \cite{Graner:2016ses}} & \multirow{2}{*}{$7.4\times10^{-30}$} & $|d_n|: 1.6\times10^{-26}$ \\
 &&$|d_p|: 2.0\times10^{-25}$\\
 \hline
 Neutron \cite{Baker:2006ts} & $4.2\times10^{-26}$ & $|d_n|: 4.2\times10^{-26}$\\
 \hline
\end{tabular}}%
\caption{The most relevant EDM experimental limits at 95\% C.L. .} 
\end{table}
The proton EDM is expected to be studied by the CPEDM Collaboration using a proton ring at CERN in the future (2021+), strongly improving upon the proton EDM limit: 
\begin{equation}
|d_p| < 2 \times 10^{-29} \;{\rm e.cm}\,.
\end{equation}

\subsubsection{EDMs in the MSSM}

In the MSSM, the EDMs are affected by different sectors of the theory, such that \cite{Ellis:2008zy}:
\begin{equation}
d_f=d_f^{\tilde{\chi}^\pm}+d_f^{\tilde{\chi}^0}+d_f^{\tilde{g}}\;(+ {\rm{higher\; order}}\;d_f^{H})\, . 
\end{equation}
where $f=e,\mu,u,d,s$. The chargino-mediated one-loop EDMs are given by:
\begin{eqnarray*}
 d_l^{\tilde{\chi}^\pm} &=& - \frac{e}{16\pi^2} \sum_i \frac{m_{\tilde{\chi}_i^\pm}}{m^2_{\tilde{\nu}_l}}
 \mbox{Im} \left( g_{Ri}^{\tilde{\chi}^\pm l\tilde{\nu}*} g_{Li}^{\tilde{\chi}^\pm l\tilde{\nu}} \right) f(m^2_{\tilde{\chi}_i^\pm}/m^2_{\tilde{\nu}_l})\,, \\
 d_u^{\tilde{\chi}^\pm} &=& \frac{e}{16\pi^2} \sum_{i,j} \frac{m_{\tilde{\chi}_i^\pm}}{m^2_{\tilde{d}_j}}
 \mbox{Im} \left( g_{Rij}^{\tilde{\chi}^\pm u\tilde{d}*} g_{Lij}^{\tilde{\chi}^\pm u\tilde{d}} \right)\,,\\ 
 &&\times\Bigl[f(m^2_{\tilde{\chi}_i^\pm}/m^2_{\tilde{d}_j}) - \frac13 g(m^2_{\tilde{\chi}_i^\pm}/m^2_{\tilde{d}_j})\Bigr] \,,\\
 d_d^{\tilde{\chi}^\pm} &=& \frac{e}{16\pi^2} \sum_{i,j} \frac{m_{\tilde{\chi}_i^\pm}}{m^2_{\tilde{u}_j}}
 \mbox{Im} \left( g_{Rij}^{\tilde{\chi}^\pm d\tilde{u}*} g_{Lij}^{\tilde{\chi}^\pm d\tilde{u}} \right)\\ 
 &&\times\Bigl[-f(m^2_{\tilde{\chi}_i^\pm}/m^2_{\tilde{u}_j}) + \frac23 g(m^2_{\tilde{\chi}_i^\pm}/m^2_{\tilde{u}_j})\Bigr]\,.
\end{eqnarray*}
The neutralino-mediated one-loop EDMs are:
\begin{equation}
 d_f^{\tilde{\chi}^0} = \frac{e}{16\pi^2} \sum_{i,j} \frac{m_{\tilde{\chi}_i^0}}{m^2_{\tilde{f}_j}}
 \mbox{Im} \left( g_{Rij}^{\tilde{\chi}^0 f\tilde{f}*} g_{Lij}^{\tilde{\chi}^0 f\tilde{f}} \right) Q_{\tilde{f}} \,g(m^2_{\tilde{\chi}_i^0}/m^2_{\tilde{f}_j})\,.
\end{equation}
The gluino-mediated one-loop EDMs read:
\begin{equation}
 d_q^{\tilde{g}} = \frac{e}{3\pi^2} \sum_i \frac{m_{\tilde{g}}}{m^2_{\tilde{q}_i}}
 \mbox{Im} \left( g_{Ri}^{\tilde{g} q\tilde{q}*} g_{Li}^{\tilde{g} q\tilde{q}} \right) Q_{\tilde{f}}\, g(m^2_{\tilde{g}}/m^2_{\tilde{q}_i})\,.
\end{equation}
The $g^{\tilde{G}ff'}$ contains the gaugino/neutralino/squark mixing matrices, and $Q_{\tilde{f}}$ is the charge of $\tilde{f}$.
The remaining term $d_f^{H}$ is affected by the Higgs bosons and corresponds to higher order terms.
The EDMs are therefore sensitive to the CP-violating phases as well as the masses of the sfermions and gauginos.

\subsubsection{Geometric approach}

\begin{figure}[!t]
\begin{center}
\includegraphics[width=5.5cm]{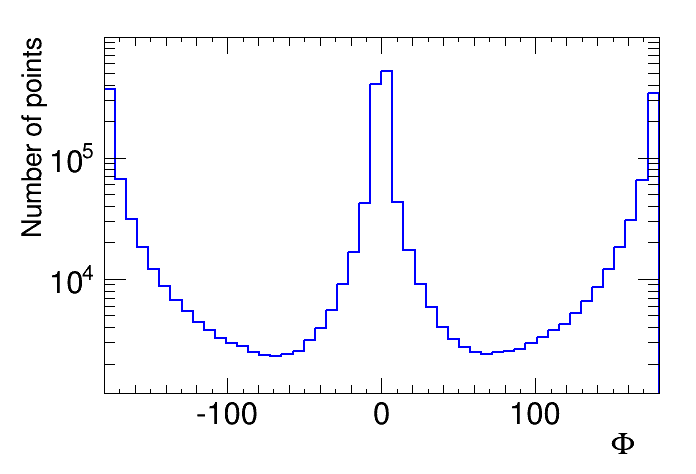}
\caption{Distributions of the phases prior to imposing EDM constraints.\label{fig:initphases}}
\end{center}
\end{figure}
\begin{figure}[!t]
\begin{center}
\includegraphics[width=4.cm]{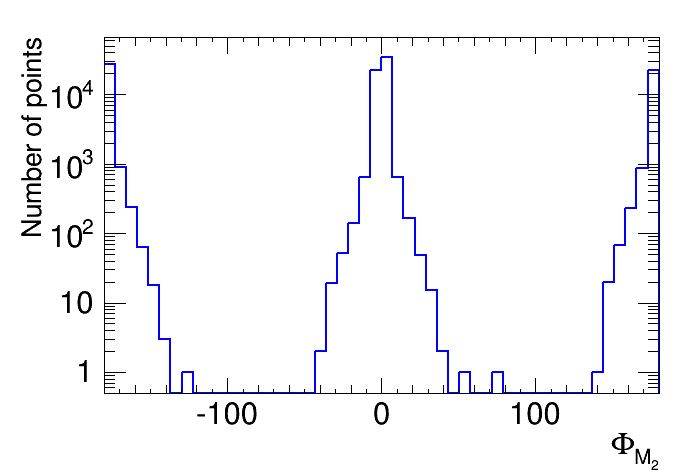}\includegraphics[width=4.cm]{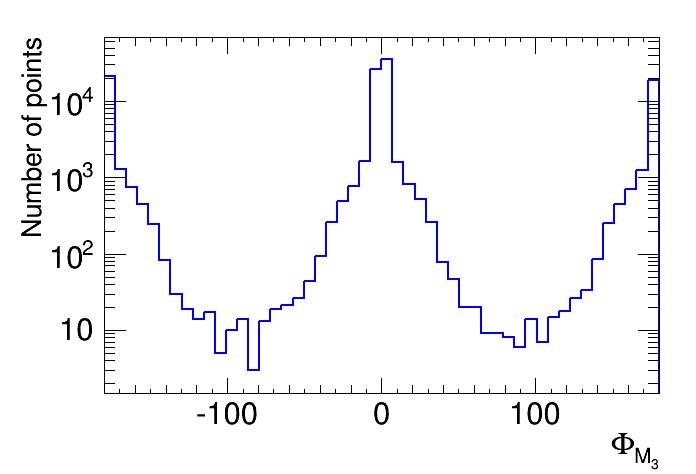}\\
\includegraphics[width=4.cm]{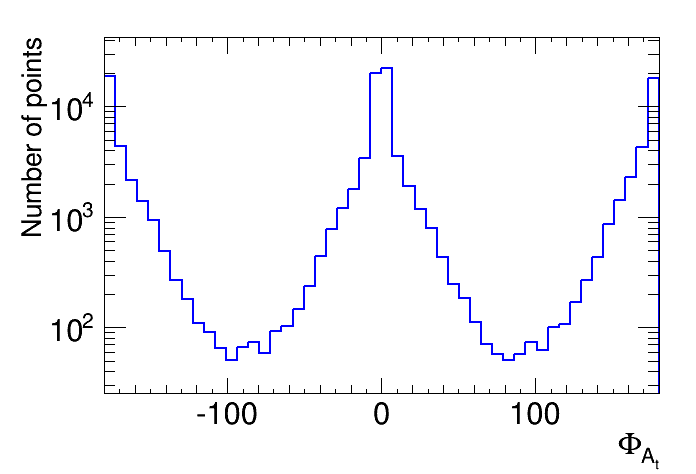}\includegraphics[width=4.cm]{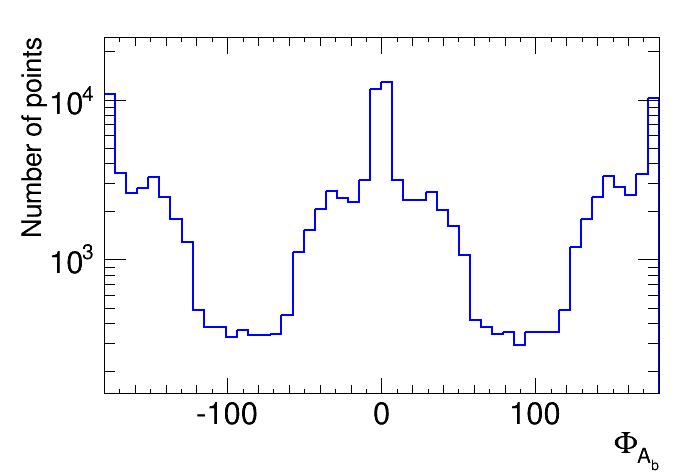}
\caption{Distributions of the $\Phi_{M_2}$, $\Phi_{M_3}$, $\Phi_{A_t}$ and $\Phi_{A_b}$ phases after imposing the current EDM constraints.\label{fig:currentphases}}
\end{center}
\end{figure}

%
A difficulty appears when applying the EDM constraints to the CP-violating pMSSM parameter space: the thorium monoxyde imposes a limit so strong that only zero phases would pass the constraints in a random scan, due to the limited statistics in a 25 dimensional parameter space. To access the regions where large phases are still compatible with the experimental results, which generally correspond to degeneracies between the different components of the EDMs, we use the geometric approach described in Refs.~\cite{Ellis:2010xm,Arbey:2014msa,Arbey:2016cey}. The idea is to determine the direction in the phase parameter space minimising the EDMs, $E^ i$, and maximising another CP-violating observable, $O$. We showed that the optimal direction, computed for each choice of the 19 CP-conserving pMSSM parameters, is given by: 
\begin{equation} 
 \phi^*_\alpha = \epsilon_{\alpha\beta\gamma\delta\mu\eta} \, \epsilon_{\eta\nu\lambda\rho\sigma\tau} \, E^a_\beta \, E^b_\gamma \, E^c_\delta \, E^d_\mu \, O_\nu \, E^a_\lambda \, E^b_\rho \, E^c_\sigma \, E^d_\tau \,,
\end{equation}
with $\phi_\alpha = \phi_{1,2,3,t,b,\tau}$, $E^i_\alpha \equiv \partial E^i/\partial \phi_\alpha$ and $O_\alpha \equiv \partial O/\partial \phi_\alpha$.
The obtained direction remains correct in the limit of small phases, and we use an iterative approach to reach larger phases: we start with phases at 0, determine the optimal direction, move by at most 20 degrees, and iterate to determine the optimal direction at the new position. After imposing in addition flavour constraints, cosmological upper bound on the dark matter density, direct detection limit and requiring squarks and gluinos to have masses above 500 GeV, the obtained distribution of phases is given in Figure~\ref{fig:initphases}, and is similar for all phases. After imposing the current EDM constraints, the distribution is modified differently for each of the CP phases. The four most affected ones are shown in Figure~\ref{fig:currentphases}.

First, after imposing the EDM constraints the statistics is strongly reduced, by a factor 30. Whereas $\Phi_{M_3}$, $\Phi_{A_t}$ have shapes still similar to the original distribution, the distribution of $\Phi_{A_b}$ is deformed preferring intermediate values of $\Phi_{A_b}$, and $\Phi_{M_2}$ is strongly affected showing that the weakly-interacting sector is severely constrained. This is mainly led by the thorium monoxide EDM limit, which sets a very strong constraint on the electron EDM.
\begin{figure}[!t]
\begin{center}
\includegraphics[width=4.cm]{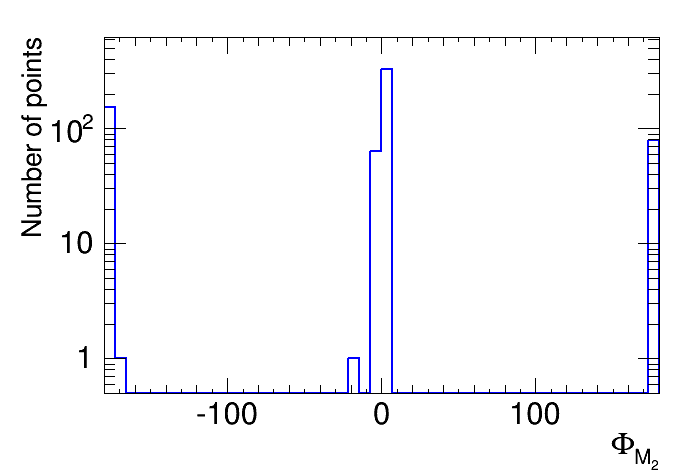}\includegraphics[width=4.cm]{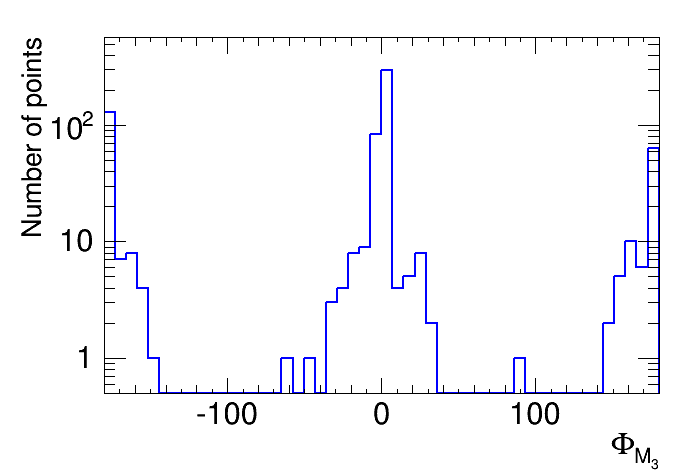}
\caption{Distributions of the $\Phi_{M_2}$ and $\Phi_{M_3}$ phases after imposing the current EDM constraints and the prospective proton EDM limit.\label{fig:futurephases}}
\end{center}
\end{figure}%

On the contrary, phases from the strongly interacting sector are less affected, because of the much weaker nucleon EDM limits. The CPEDM prospect to measure precisely the proton EDM will therefore be extremely useful. Figure~\ref{fig:futurephases} shows the distribtuions of the phases after imposing a limit on the proton EDM of $2\times10^{-29}$ e.cm. As can be seen, the statitics is again strongly reduced, and $\Phi_{M_2}$ is restricted to nearly zero phases. In addition, $\Phi_{M_3}$ becomes strongly affected, and only phases up to $30^{\circ}$ could survive. This shows that the strongly interacting sector can be tested, providing a complementary way to probe the gluino and squark masses and couplings along with the SUSY direct searches at the LHC, in the case additional sources of CP-violation exist.

\subsubsection{Other CP-violation sensitive observables}

The SUSY direct searches are not affected by CP-violation, even if the sparticle spectrum can be. The Higgs sector can on the contrary be sensitive to CP-violation. However, the 125 GeV Higgs has been measured to be CP-conserving, severely limiting the possibility to have a pseudoscalar component. Consequently, only heavier Higgs states could reveal CP-violation. The observations of the decay of heavy Higgs states into tops or taus of specific polarisations would help probing the CP-violating content of the Higgs bosons \cite{Chakraborty:2013si}. Concerning the dark matter sector, we showed that it is rather insensitive to CP-violation \cite{Arbey:2014msa}.

\begin{figure}[!t]
\begin{center}
\includegraphics[width=7cm]{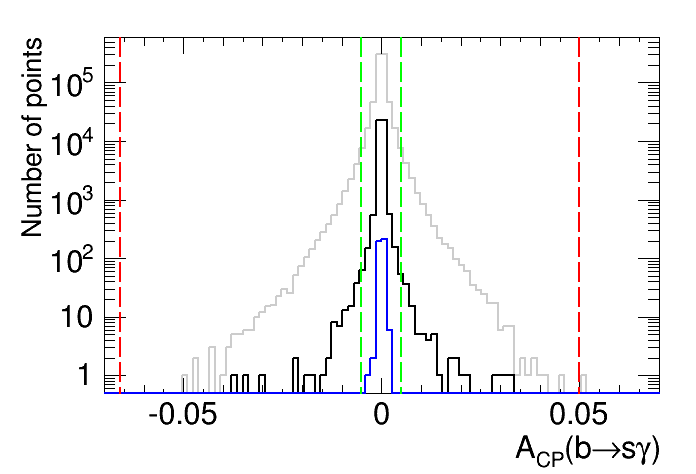}
\caption{Distribution of the CP asymmetry of $b \to s \gamma$. The gray curve corresponds to the number of points in absence of EDMs, the black curve using the current EDM limits and the blue one also considering the prospective proton EDM. The red lines correspond to the current $b \to s \gamma$ CP asymmetry limits and the green one to the prospective limits of Belle-II.\label{fig:bsgamma}}
\end{center}
\end{figure}

Flavour physics on the other hand is very sensitive to CP-violation. Two observables are particularly relevant for our study: the CP asymmetry in $b \to s \gamma$ and the $B_s$ meson mixing $\Delta M^{NP}_{B_s}$. In the MSSM, $b\to s\gamma$ is sensitive to the charged Higgs, gaugino and squark sectors. Charged Higgs is mostly insensitive to CP-violation, but because of chargino/stop loops CP-violation can be visible in $b \to s \gamma$. CP asymmetry in $b \to s \gamma$ has been measured at B factories, but the current limits are not strong enough to provide insightful constraints on CP-violation in the pMSSM. Belle-II will however provide in the near future stronger limits~\cite{Kou:2018nap}. Figure \ref{fig:bsgamma} shows the distribution of CP asymmetry in $b \to s \gamma$ depending on the EDM constraints, as well as the current and future experimental limits. We note that the current EDM limits supersede the measurements of CP asymmetry. The future Belle-II results will strongly improve on the limits, and probe CP-violation in the MSSM. It is remarkable to see that the future proton EDM measurement will provide comparable constraints on CP-violation.

\begin{figure}[!t]
\begin{center}
\includegraphics[width=7cm]{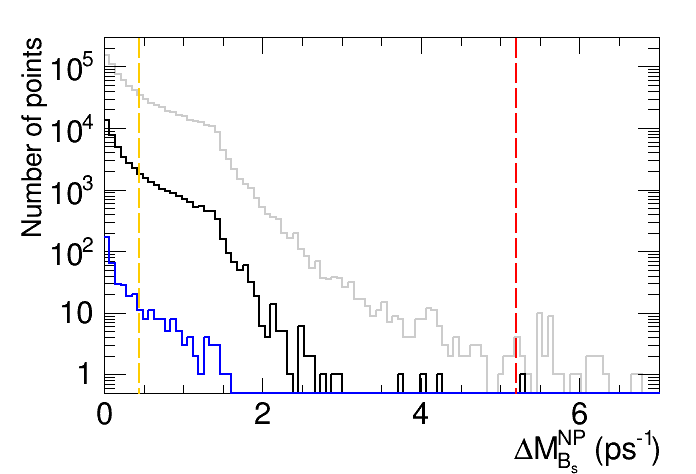}
\caption{Distribution of the $\Delta M^{NP}_{B_s}$. The gray curve corresponds to the number of points in absence of EDMs, the black curve using the current EDM limits and the blue one also considering the prospective proton EDM. The red line corresponds to the current theoretical uncertainties and the yellow one to the prospective uncertainties.\label{fig:deltaMBs}}\vspace*{-0.2cm}
\end{center}
\end{figure}

Similarly the $B_s$ meson mixing is strongly sensitive to the gaugino and squark sectors \cite{Gabbiani:1996hi}. However, the current experimental measurements are extremely precise, and the main limitation comes from the theoretical uncertainties from the determination of form factors. We can nevertheless expect in the future an improvement by a factor ten on the uncertainties. In Figure \ref{fig:deltaMBs}, the distributions of $\Delta M^{NP}_{B_s}$ depending on the EDM constraints is shown, as well as the current and future limits including the theoretical uncertainties. The current limits are not competitive with the EDM constraints for CP-violation, but a strong reduction of the theoretical uncertainties would allow for a significant improvement in probing CP-violation.

\section{Conclusions}

We have discussed the interplay between the contraints from different sectors to probe the MSSM parameter space. We have shown that it is important to combine the constraints from direct searches for Higgs and supersymmetric particles at the LHC, flavour physics and dark matter observables to study the 19-parameter space of the phenomenological MSSM. In presence of additional sources of CP-violation, we have also quantified the importance of electric dipole moments and flavour observables sensitive to CP-violation. We have shown that a measurement of the proton EDM as projected by the CPEDM collaboration is of utmost importance, and will be complementary to the theoretical and experimental improvements in flavour physics in order to deeply probe CP-violation.

\section*{Acknowledgements}
The author is grateful to the organisers of the Capri workshop for their invitation and the great workshop, and to A. Arbey, M. Battaglia, J. Ellis and G. Robbins for their collaboration on the material presented here.

\nocite{*}
\bibliographystyle{elsarticle-num}



\end{document}